\documentclass[twocolumn,showpacs,prl]{revtex4}

\usepackage{epsfig,latexsym,graphicx,amsfonts,bm,pifont}
\newcommand{\beq}{ \begin{equation} }
\newcommand{\eeq}{ \end{equation} }
\newcommand{\beqs}{ \begin{eqnarray} }
\newcommand{\eeqs}{ \end{eqnarray} }

\newcommand{\f}[2]{\frac{#1}{#2}}
\newcommand{\vect}[1]{\ensuremath{%
        {\mbox{\mathversion{bold}\ensuremath{#1}}}%
        }}

\begin{document}

\title{Rotational dynamics of a superhelix towed in a Stokes fluid}% Force line breaks with \\

\author{Sunghwan Jung$^1$}
\author{Kathleen Mareck$^{1,2}$}
\author{Lisa Fauci$^{2}$}
\author{Michael J. Shelley$^1$}
\affiliation{$^1$Applied Mathematics
Laboratory, Courant Institute of Mathematical Sciences, New York
University,
251 Mercer Street, New York, New York 10012, USA \\
$^2$Department of Mathematics, Tulane University, 6823 St. Charles
Avenue, New Orleans, Louisiana 70118, USA}

\date{\today}

\begin{abstract}
Motivated by the intriguing motility of spirochetes (helically-shaped
bacteria that screw through viscous fluids due to the action of
internal periplasmic flagella), we examine the fundamental fluid
dynamics of superhelices translating and rotating in a Stokes fluid.
A superhelical structure may be thought of as a helix whose axial
centerline is not straight, but also a helix.  We examine the
particular case where these two superimposed helices have different
handedness, and employ a combination of experimental, analytic, and
computational methods to determine the rotational velocity of
superhelical bodies being towed through a very viscous fluid.  We find
that the direction and rate of the rotation of the body is a result of
competition between the two superimposed helices; for small axial
helix amplitude, the body dynamics is controlled by the short-pitched
helix, while there is a cross-over at larger amplitude to control by
the axial helix.  We find far better, and excellent, agreement of our
experimental results with numerical computations based upon the method
of Regularized Stokeslets than upon the predictions of classical
resistive force theory.
\end{abstract}

%\pacs{Put Valid PACS}% PACS, the Physics and Astronomy
\maketitle

\section{Introduction}

The study of swimming micro-organisms, including bacteria, has long
been of scientific interest \cite{Taylor52,Berg73,Lighthill76}.
Bacteria swim by the action of rotating, helical flagella driven by
reversible rotary motors embedded in the cell wall \cite {Berg73}.
Typically, these flagella visibly emanate from the cell body.  The
external flagella of rod-shaped bacteria, such as {\em E. coli}, form
a coherent helical bundle when rotating counter-clockwise, causing
forward swimming.  When these flagella rotate in the opposite
direction, the flagellar bundle unravels, causing the cell to tumble.
This run and tumble mechanism allows a bacterium to swim up a
chemo-attractant gradient as it senses temporal changes in
concentration \cite{Brownberg, Macnab}.  Many studies have focused on
the fundamental fluid mechanics surrounding this locomotion affected
by a simple helical flagellum attached to and extruded from the cell
body \cite{Lighthill76, Purcell97}.  Recently, there have been
additional studies that investigate the hydrodynamics of flagellar
bundling \cite{Powers, Flores}.

In contrast, swimming bacteria with more complicated body-flagella
arrangements are less studied.  Spirochetes are such a group of
bacteria. They have a helically-shaped cell body \cite {CDMF84,Lepto},
and although they also swim due to the action of rotating flagella,
these do not visibly project outward from their cell body.  Instead,
the cell body is surrounded by an outer sheath, and it is within this
periplasmic space that rotation of periplasmic flagella (PFs) occurs.
These helical periplasmic flagella emanate from each end of the cell
body, but rather than extend outwards, they wrap back around the helical
cell body.  In the case of {\em Leptospiracaeae}, there are two PFs,
one emerging from each end of the cell body, that do not overlap in
the center of the cell.  Rotation of each flagella is achieved by a
rotary motor embedded in the cell body.  The shapes of both ends of
the helical cell body are then determined by the intrinsic helical
structure of the periplasmic flagella, as well as their direction of
rotation.  During forward swimming, {\em L. illini} exhibit an
anterior region that is superhelical, due to this interplay of helical
cell body and helical flagellum \cite {Lepto}.  In fact, the
handedness of these two helical structures are opposite, with the
flagellum (axial helix) exhibiting a much larger pitch than the cell
body helix.

The overall swimming dynamics of spirochetes involves non-steady
coupling of the complex geometry of the cell body, the flexible outer
sheath, and the counter-rotation of the cell body with the internal
flagella.  However, a natural question is how the effectiveness of
spirochete locomotion depends upon the detailed superhelical geometry
of the anterior region of the bacterium.  With this as motivation, we
present here a careful study of the fundamental fluid mechanics of
superhelical bodies translating and rotating through highly viscous
fluids.  We extend the classical analytic and experimental results of
Purcell \cite {Purcell97} and the numerical results of Cortez et
al. \cite {Cortez05} performed for regular helices.  In addition, we
offer coordinated laboratory and computational experiments as
validation of the method of Regularized Stokeslets for zero Reynolds
number flow coupled with an immersed, geometrically complex body.
This method uses modified expressions for the Stokeslet in which the
singularity has been mollified. The regularized expression is derived
as the exact solution to the Stokes equations consistent with forces
given by regularized delta functions.

We focus on a typical body that is a short-pitched helix whose axis is
itself shaped as a helix of larger pitch and opposite handedness.  In
the following sections, we describe the experimental set-up as well as
the construction of these superhelical bodies.  We experimentally
measure the rotational velocities of the bodies as they are towed with
a constant translational velocity through a very viscous fluid.  Note
that rotational and translational velocities should be proportional,
with the constant of proportionality (resistance coefficient)
dependent upon the body geometry.  The rotational velocities
corresponding to translational velocities are also predicted
analytically using resistive force theory, as well as using the method
of Regularized Stokeslets \cite {Cortez01, Cortez05}.  We find
compatible behavior between experiments and the resistive force
theory, but excellent quantitative agreement between experiments and
the method of Regularized Stokeslets.

\section {Superhelix construction}

\begin{figure} %%Exp_HE.ai
    \centering
        \includegraphics[width=.5\textwidth]{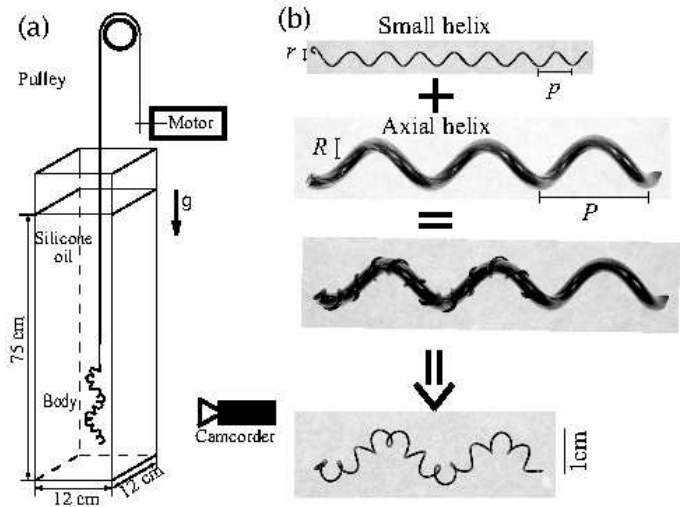}
\caption{(a) Schematic of experimental setup. A motor pulls a
rigid body through silicon oil, a highly viscous Stokes fluid
($\nu$ = 10,000 cSt) (b) Procedure for making superhelix.}
    \label{Figure_Exp}
\end{figure}

A superhelix is formed from a copper wire chosen to be sufficiently
malleable to deform into a desired shape, but rigid enough not to
deform as it moves through the viscous fluid.  The superhelix is made
in two steps (see Fig.\,\ref{Figure_Exp}(b)). First, a copper wire of
diameter 0.55 mm is wound tightly in a clockwise direction up a rod of
diameter 3.15 mm, forming a tight coil.  After removing the coil from
the rod, we stretch it out into a smaller radius, larger pitch helix, simply by
pulling the ends of the coil away from each other. The axial helix is
made in the same manner, but we use lead wire of a thicker diameter
(3.15 mm) and a larger rod (4.7 mm diameter). The most important
difference between the two helices is handedness; the axial helix is
wound counter-clockwise up the rod, whereas the small helix is wound
clockwise. Once the parameters of the small and axial helices are
measured, the axial helix is threaded through the small helix, forming
a superhelix, i.e. the small helix is placed back on a rod that has
been distorted into a helical shape. The last step is to remove the
axial helix. This is done by simply rotating the axial helix while
keeping the superhelix fixed.

%%%%%%%%%%%%%%%%%%%%%%%%%%%%%%%%%%%%%%%%%%%%%%%%%%%%%%%%%%%%%%%
%%%%%%%%%%%%%%%%%%%%%%%%%%%%%%%%%%%%%%%%%%%%%%%%%%%%%%%%%%%%%%%%%

%% Type I=> G Type II => H

\begin{figure} %%Plot_Radius_Pitch.m
    \centering
        \includegraphics[width=.5\textwidth]{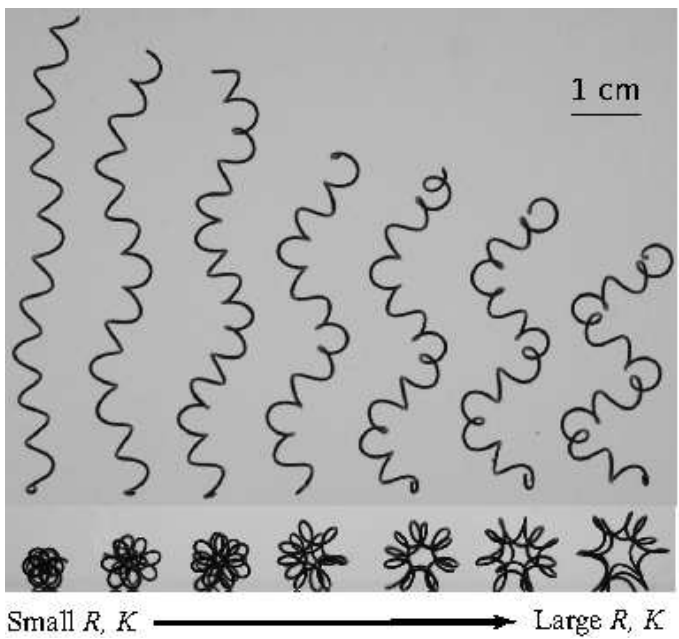}
\caption{
Seven superhelices with increasing axial helix radius (from left to right).
The upper panel is the side view and the lower panel is the axial view.
%Radius ($R$) and wavenumber ($K$) of the axial helix. As
%the initial coil is pulled apart, the radius and wavenumber of the
%axial helix increase. This relation can be predicted by the simple
%scaling relation with inextensibility of a wire. 
}
    \label{Figure_RadPit}
\end{figure}

The defining geometric parameters of the superhelix are the radius $r$
and pitch $p$ of the small helix, and the radius $R$ and pitch $P$ of
the axial helix (see Fig.\,\ref{Figure_Exp}).  For $R=0$ or $P=\infty
$, the superhelix reduces to a regular helix.  In our experiments, two
different sets of small helices are used.  The corresponding geometric
parameters of these small helices are the pitch ($5.58 \pm 0.25$ mm;
Set I and $5.04 \pm 0.36$ mm; Set II) and radius ($1.91 \pm 0.14$ mm;
Set I and $1.75 \pm 0.21$ mm; Set II).  The small (less than 12\%)
variations of pitch and radius are presumably due to mechanical
relaxation of material when it is pulled off the axial helix.  Seven
different axial helices are prepared from the same initial coil (see
Fig. \ref{Figure_RadPit}).

%As the two ends of coil are pulled
%apart, the radius of the resulting axial helix decreasesm and the pitch
%increases.
%A body structure of superhelices can
%be considered as a superposition of coordinates of two helices.

We now construct a mathematical representation of the superhelix.  The
coordinates of an axial helix are $ \vect{ X} = (R \cos (Kz) , R \sin
(Kz), z)$, where $K = \frac{2 \pi}{R}$. The distance measured along
this helix is linearly proportional to the axial distance $z (= \alpha
s)$. The unit vector tangential to the axial helix is
$\hat{\vect{t}}_A = {\partial \vect{X}}/{ \partial s}$. The principal
normal vector is $\hat{\vect{n}}_A =
%(1/\kappa)\partial \hat{\vect{t}}_A/
%\partial s_A = 
(-\cos(Kz), -\sin(Kz), 0)$ and the binormal is
$\hat{\vect{b}}_A = \hat{\vect{t}}_A \times \hat{\vect{n}}_A =
\alpha(\sin(Kz),-\cos(Kz),RK)$. 
%Inextensibility condition of the
%helix requires
Since $\hat{\vect{t}}_A$ is a unit vector sets $\alpha$ as:
\beqs
\alpha^2(R^2 K^2 +1) &=& 1 \,.
\eeqs
%The construction of the axial helix from a coil with a given arclength
%is done by stretching the axial distance $z$ 
%by $\lambda$. In order to keep the same arclength,
%$\alpha \rightarrow \lambda \alpha$ and $K \rightarrow (1/\lambda)
%K$. From those two equations, one gets $R \rightarrow R \sqrt{1 +
%(1-\lambda^2)/R^2 K^2 }$. It shows that $R^2 + 1/K^2 = 1/C^2$,
%where $C$ is a constant. This relation with $C=0.24$ (1/mm) is
%plotted in Fig. \ref{Figure_RadPit}, along with the corresponding
%measurements of our constructed helices. 

The coordinates of the one-dimensional curve describing the superhelix are:
\begin{eqnarray}
&&{\vect R}(s) = (R_x, R_y, R_z) \nonumber \\
&&= {\vect X}(s) + r \cos (ks) \hat{{\vect n}}_A + r \sin
(ks) \hat{\vect{b}}_A \,.
\end{eqnarray}
\noindent
Recall that the actual superhelices have nonzero thickness (the diameter of the copper wire), and
hence are true three-dimensional structures.
%Fig. \ref{Figure_RadPit} shows the constructed helices
%from both a sideview and an axial view.  

%Fig. \ref{Figure_RadPit}. 
%As a superhelix is close to a small
%helix ($R \rightarrow 0$ or $P_L \rightarrow \infty$), the
%non-dimensional number $RK$ approaches to zero.

%%%%%%%%%%%%%%%%%%%%%%%%%%%%%%%%%%%%%%%%%%%%%%%%%%%%%%%%%%%%%%
%%%%%%%%%%%%%%%%%%%%%%%%%%%%%%%%%%%%%%%%%%%%%%%%%%%%%%%%%%%%%%%%

\section {Experiment}

The classical experiments of Purcell, elaborated on in \cite{Purcell97},
examined the relationship between angular and translational velocities of
helical objects at very low Reynolds numbers.  
Here we extend these experiments to the superhelical objects described above.
The experimental setup was originally  designed for
sedimentation experiments \cite{Jung06} (see Fig.
\ref{Figure_Exp}(a)). A tall transparent container is filled with
silicone oil with large viscosity ($\nu=10^4$ cS, $\rho=0.98$
g/cm$^3$). The oil behaves as a Newtonian fluid in the regime of
interest here. 
Rather than allowing the superhelical object to descend by gravity, our experiment
is designed to measure its rotational speed as it is towed up through the viscous
column of fluid at a specified translational speed.  
To drag the superhelix,
a small hook ($\sim$ 2 mm) is used to attach the superhelix
to a thread from a motor. Note that the dimensions of this hook are quite small compared to the superhelix length
($\sim 4$ cm). 
By experimentally testing with an axisymmetric body (sphere), we
found that this towing system (the thread plus the motor), does not
produce any torque on the body.

\begin{figure} %%figure_linearplot.m
    \centering
        \includegraphics[width=.5\textwidth]{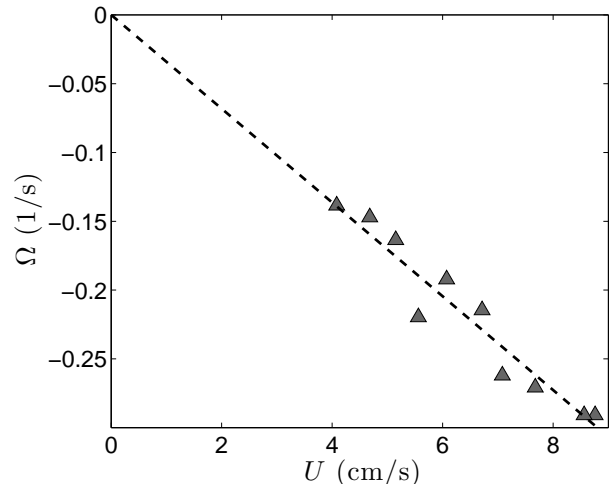}
\caption{A linear relationship between the translational velocity
and rotational frequency of a superhelix ($r = 0.89$ mm, $R =
4.62$ mm, $p = 5.5$ mm, and $P = 19.4$ mm). Triangles are from
experimental observations. Dashed line is a least-square fit of
experimental data.}
    \label{Figure_linear}
\end{figure}

The superhelix is initially positioned near the bottom of the container, then
is dragged upwards by the motor (Clifton Precision-North) at   constant
speed. In the intermediate region in the container, steady
state motion (constant translational velocity, rotational velocity, and
drag force) is assumed. 
The superhelix positions, orientations, and
velocities are measured from a 30 frames per second video stream
of the camcorder.
%The effects of the boundary and the free
%surface are not taken into account in this study, and are believed to be small.
The translational velocity in our experiments varies by changing power
input to the motor.  We have chosen a velocity range of 3-10
cm/s. Below 3 cm/s, the step motor produces non-uniform pulsed axle
rotations, which lead to irregular translational velocity.  The
Reynolds number based upon the towing velocity and radius of the
superhelical structure (1 cm) is at most:
\beq
Re = \f{U R}{\nu} \sim 0.1 \,.
\eeq
%Our experiments are presumed to be in a regime of low Reynolds
%numbers. 
We assume therefore that the steady Stokes equations 
govern the fluid mechanics of the translating superhelix.
Within this translational
velocity range, a linear relationship between rotational
velocity $\Omega$ and translational velocity $U$ is observed (see Fig. \ref {Figure_linear}).

\begin{figure} %%result_tes.m
    \centering
        \includegraphics[width=.5\textwidth]{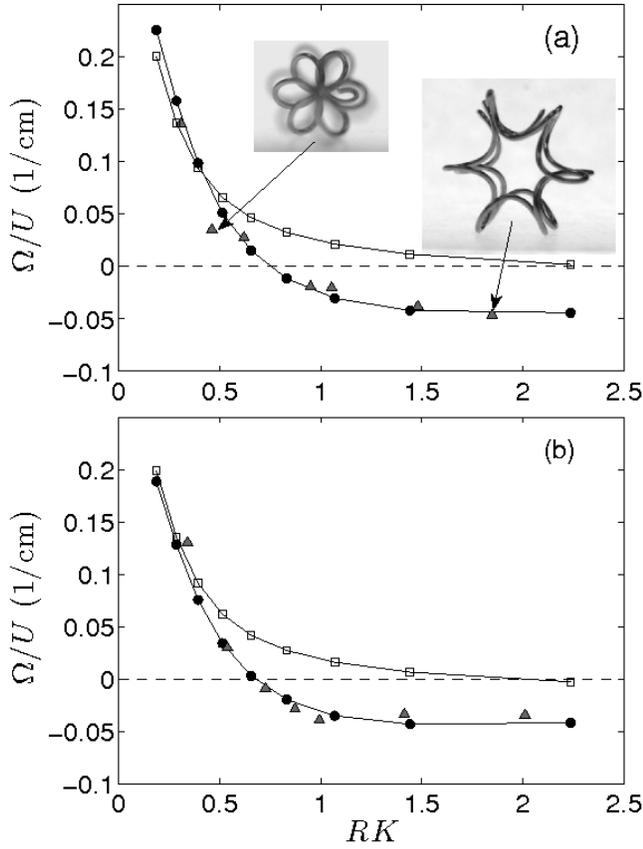}
\caption{Ratio of angular velocity to translational towing velocity.
Triangles are values measured experimentally.
Circles
connected by lines are values predicted using the method of Regularized Stokeslets. 
Squares are values predicted using  resistive force theory. 
(a) Superhelices of Set I. Positive rotational
rate is clockwise and negative rate is counter-clockwise.
Rotational direction changes around 0.7 for $RK$. (b) Superhelices
for Set II. Same transition also occurs around 0.7 for $RK$.}
    \label{Figure_transition}
\end{figure}

A translating helix in a viscous solution rotates in the direction
which it screws in. Following this rule, the small (straight) helix in
our experiments would rotate clockwise and the axial (straight) helix
would rotate counter-clockwise when viewed from above.  In Purcell's
work \cite{Purcell97}, the jointed structure built by connecting two
helices of opposite handedness, otherwise identical, showed no
rotation during its sedimentation.  The superhelix of interest here is
the superposition of two helices with opposite handedness.  The
inherent rotational directions of these superimposed helices are in
competition.  For very small values of the non-dimensional parameter
$RK$ of the axial helix, the superhelical structure reverts to the
straight small helix, and would rotate clockwise.  One expects that
for larger values of the parameter $RK$, the axial helix would be
dominant, and the superhelical structure would rotate
counter-clockwise.  For some critical value of $RK$, we would expect a
transition in direction, and hence, a structure that would show no
rotation as it is towed through the fluid.  We performed experiments
that systematically varied $RK$, and observed this expected change in
rotational direction.  Fig. \ref{Figure_transition} shows the ratio of
angular velocity to translational velocity as a function of $RK$, for
the two different sets of superhelices (Sets I \& II).  Positive
rotational rate is clockwise, and negative is counter-clockwise.  In
each set of experiments, the measured ratio is depicted by triangles.
In the next sections, we describe mathematical formulations that model
these observations.

%Resistive force theory has been proposed by
%Hancock\cite{Hancock53} and Gray and Hancock\cite{Gray55} and is
%extended by Chwang and Wu\cite{Chwang71} and Lighthills
%\cite{Lighthill76}.

%%%%%%%%%%%%%%%%%%%%%%%%%%%%%%%%%%%%%%%%%%%%%%%%%%%%%%%%%%%%%%%
%%%%%%%%%%%%%%%%%%%%%%%%%%%%%%%%%%%%%%%%%%%%%%%%%%%%%%%%%%%%%%%%%

%%%%%%%%%%%%%%%%%%%%%%%%%%%%%%%%%%%%%%%%%%%%%%%%%%%%%%%%%%%%%%%
%%%%%%%%%%%%%%%%%%%%%%%%%%%%%%%%%%%%%%%%%%%%%%%%%%%%%%%%%%%%%%%%%

%where
%\begin{eqnarray}
% &&R_x= (R -r\cos(ks_A))\cos(K\alpha s_A)-r\alpha \sin(ks_A)
%\sin(K\alpha s_A) , \nonumber \\
%&&R_y = (R -r\cos(ks_A))\sin(K\alpha s_A)+r\alpha \sin(ks_A)
%\cos(K\alpha s_A) , \nonumber \\
%&&R_z = \alpha s_A - \alpha rRK \sin(ks_A)\,,
%\end{eqnarray}
%where $k = 2\pi/ P_s$ ($P_s$ is the pitch of a small helix) and
%$r$ is the radius of small helix.

\section{Numerical Results}
\subsection {Regularized Stokeslets}

We assume that the superhelix is a rigid body moving in a Stokes fluid. The
governing equations of motion are:
\beq
- {\vect \nabla} p + \mu \nabla^2 {\vect u} = 0, ~ \nabla
\cdot \vect{u} =0 \,.
\eeq
%These equations hold in all of three-space, and $\vect{f}$ is the force of the superhelix on
%the fluid, and is a Dirac delta-function layer supported
%by  the material points of the superhelix surface. 
%Integrating over the surface $\partial D$ of the superhelix gives us the 
The total hydrodynamic force and torque exerted by the superhelix (with surface $\partial D$) on the surrounding fluid is:
\begin{equation}
\begin{array}{rcl}
{\bf F} &= &\int\limits_{{\bf x} \in \partial D} {{\bf f}({\bf x}) d {\bf x}},
\end{array}
\label{totalforce}
\end{equation}
\begin{equation}
\begin{array}{rcl}
{\bf L} &= &\int\limits_{{\bf x} \in \partial D} {({\bf x}-{\bf x_0}) \times {\bf f}({\bf x}) d {\bf x}}.
\end{array}
\label{totalmoment}
\end{equation}

\noindent
where ${\bf f}$ is the surface traction.

A solution to the Stokes equations in 3D  with a point force centered at $\vect{x}_0$ is
the classical Stokeslet \cite {Poz}.  Due to the linearity of the Stokes equations, 
superposition of these fundamental solutions allows the
construction of the velocity field induced by a distribution of  point forces.  
The method of 
Regularized Stokeslets eases the evaluation of integrals with singular kernels
by replacing the delta distribution of forces by a smooth, localized distribution
\cite{Cortez01,Cortez05}. 
The  force  $\vect{f} = \vect{f}(\vect{x}_0) \delta (\vect{x} - \vect{x}_0)$ is replaced 
by $\vect{f} = \vect{f} 
(\vect{x}_0) 
\phi_{\epsilon}
(\vect{x} - \vect{x}_0)$
where 
$\phi_{\epsilon}$ is a cutoff, or blob, function with integral one.  
This blob function is an approximation to the 3D Dirac delta function, with $\epsilon$  a small parameter.
Following \cite {Cortez05}, we choose 
\beq
\phi_{\epsilon} (\vect{x} - \vect{x}_0) = \f{15 \epsilon^4}{8 \pi
(\| \vect{x} - \vect{x}_0 \|^2 + \epsilon^2)^{7/2} } \,.
\eeq
For $N$ regularized point forces distributed on the surface of a body in rigid rotation and translation, the fluid
velocity at any point $\vect{x}$ is evaluated as
\beq
{8 \pi \mu}{u}_i (\vect{x}) = \sum_j \sum_{n=1}^N
S^{\epsilon}_{ij} (\vect{x}, \vect{x}_n) {f}_{j} (\vect{x}_n)
\label{eq:u_S_f}
\eeq
For the given cutoff function, the kernel $S$ is
\beq
S^{\epsilon}_{ij} (\vect{x}, \vect{x}_n) = \delta_{ij} \f{r^2 + 2
\epsilon^2}{(r^2 + \epsilon^2)^{3/2} } + \f{(x_i - x_{n,i}) (x_j -
x_{n,j})}{(r^2 + \epsilon^2)^{3/2}} \,
\eeq
where $r = \| \vect{x} - \vect{x}_n \|$.

Note that evaluating equation (\ref{eq:u_S_f})
at each of the $N$ points of the superhelix surface gives us a linear relation between 
the velocities and the forces 
exerted 
at these points.  
The matrix $S^{\epsilon}_{ij}$, for a given cutoff parameter $\epsilon$ depends only upon the
geometry of the superhelix.  

For a rigid body moving in a Stokes flow, there is a linear relationship
between the total hydrodynamic  force and
torque and the translational and rotational velocity of the body \cite {Purcell97}.
Following \cite {Purcell97, Cortez05}, we focus on the 
$z$-components of total hydrodynamic force $F$ and torque $L$, along 
with the $z$-component of translational velocity $U$, and
rotational velocity about the $z$-axis $\Omega$. These are related by resistance
(or propulsion) coefficients: 
\beq
\left( \begin{array}{c} F \\ L \end{array} \right) = \mu \left(
\begin{array}{cc}  A &  B \\  B &  D \end{array} \right)
\left( \begin{array}{c} U \\ \Omega \end{array} \right) \,.
\label{matrices}
\eeq
Here $A$, $B$, and $D$ depend only upon the 
geometry of the object.

In order to compute these coefficients, 
we discretize the cross sections of the copper wire using six azimuthal grid points.
We choose a cutoff parameter $\epsilon$ on the order of the distance between discrete points
(see \cite {Cortez05} for details).
At each point on this discretized superhelix, we impose a unit translational velocity and zero 
rotational velocity in the $z$-direction.  
We use the Regularized Stokeslet linear relation (\ref{eq:u_S_f}) 
to solve for the forces on the superhelix that produced this velocity.
We then evaluate the integrals for total force $F$ and total torque $L$ in equations
(\ref {totalforce}, \ref {totalmoment}) above.
Using the linear equations (\ref{matrices}), we compute the resistance coefficients $A, B$.
Similarly, we can 
compute $D$ by imposing a unit rotational velocity and zero translational velocity.

These resistance coefficients allow us to predict the ratio of
rotational velocity to translational velocity in our torque-free
experiments described above, as
\beq \label{eq:linear}
\frac {\Omega}{U} = - \f{B}{D} \,.
\eeq
Figure \ref{Figure_transition} shows the Regularized Stokelet
predictions (circles) of these ratios for both sets of superhelices.
The agreement with experimental data is excellent.  Indeed, the
transition from clockwise to counter-clockwise rotation is captured
very precisely.

\subsection{Resistive Force Theory}

Resistive force theory \cite{Hancock53,Gray55} is widely used to give an approximate
description on a slender body moving in a viscous fluid.  However, the
non-local interactions of stress along the body are not taken into
account.  To see how important this non-local interaction is, we
analytically estimate the ratio of angular to translational velocities
in this section, and compare the predictions with experiment and the
numerical calculations using regularized Stokeslets.

The Stokes drag force is proportional to its velocity as $\vect{f}
= C_t (\vect{u} \cdot \hat{t}) \hat{t} + C_n (\vect{u} \cdot \hat{n})
\hat{n}$ where $\hat{t}$ and $\hat{n}$ are tangential and normal
directions, and $C_t$ and $C_n$ are drag coefficients. The normal
direction is arbitrary, but, uniquely determined if the motion is
given.

First, we consider pure body-rotation about the $z$-axis. The body
velocity is $\vect{u} = \vect{R}_{\perp} \times \vect{\Omega} = \Omega
\, (R_y, - R_x, 0)= \Omega R_{\perp} \hat{u}$ where $\vect{R}_{\perp}
= (R_x, R_y, 0) $, ${R}_{\perp} = \| \vect{R}_{\perp}\| $, $\hat{u} =
\hat{R} \times \hat{z}$, and $\vect{\Omega} = \Omega \hat{z}$.
%Here, the normal direction $\hat{n}$ is determined by using the relations ($\hat{n} \cdot \hat{t} = 0$, $\hat{t} \cdot \hat{u} = \cos \psi$, and $\hat{n}
%\cdot \hat{u} = \sin \psi$ where $\psi$ is the angle ) associated with the tangential $\hat{t}$ and the direction of motion $\hat{u} (= \hat{R}_{\perp} \times \hat{z})$.
The angle $\psi$ between the direction of motion and the tangential of
the body is expressed by the superhelix coordinates as $\hat{u} \cdot
\hat{t} = \cos \psi = (R_y \partial_s R_x - R_x \partial_s
R_y)/R_{\perp}$ where $\hat{t} = R_s$.  Similarly, $\hat{u} \cdot
\hat{n} = \sin \psi$.

Using the vector relation ($(\vect{A} \times \vect{B}) \cdot \vect{C} = - (\vect{A} \times \vect{C}) \cdot \vect{B}$), the $z$-component torques associated with forces are
\beqs
L_z(s) &=& (\vect{R} \times \vect{f}) \cdot \hat{z}  \nonumber \\
&=& C_t (\vect{u} \cdot \hat{t}) (\vect{R} \times \hat{t}) \cdot \hat{z} + C_n (\vect{u} \cdot \hat{n}) (\vect{R} \times \hat{n}) \cdot \hat{z} \nonumber \\
&=& - C_t (\vect{u} \cdot \hat{t}) (\vect{R} \times \hat{z}) \cdot \hat{t} - C_n (\vect{u} \cdot \hat{n}) (\vect{R} \times \hat{z}) \cdot \hat{n} \nonumber \\
&=& - \Omega R_{\perp}^2 ( C_t  \cos^2 \psi + C_n \sin^2 \psi ) \,,
\eeqs
and the total torque due to its rotational motion is
\beqs
&&L_z^{(Rotation)} = \int L_z(s) ds \nonumber \\
&& = - \Omega \int R_{\perp}^2 \left[ C_t \cos^2 \psi + C_n \sin^2
\psi \right] ds \,. ~~
\eeqs
%Similarly, with the pure translation $\vect{u} = (0, 0, U)=U\hat{z}$
%on a rigid body reduces the normal direction as $\hat{n} = (\hat{z} -
%(\partial R_z/ \partial s) \, \hat{t})/(1 - (\partial R_z/\partial
%s)^2)^{1/2}$.  The $z$-component torques associated with forces are
%\beqs
%L_z(s) &=& C_t (\vect{u} \cdot \hat{t}) (\vect{R} \times \hat{t})
%\cdot \hat{z} + C_n (\vect{u} \cdot \hat{n}) (\vect{R} \times \hat{n})
%\cdot \hat{z} \nonumber \\ &=& - C_t (\vect{u} \cdot \hat{t})
%(\vect{R} \times \hat{z}) \cdot \hat{t} - C_n (\vect{u} \cdot \hat{n})
%(\vect{R} \times \hat{z}) \cdot \hat{n} \nonumber \\ &=& - U C_t
%(\partial R_z / \partial s)R_{\perp} \cos \psi + U C_n (\partial R_z/
%\partial s)R_{\perp} \cos \psi \nonumber \\ &=& U ( C_n - C_t )
%(\partial_s R_z) R_{\perp} \cos \psi \,,
%\eeqs
%%$u \cdot\hat{n} = U (1 - (\partial_s R_z)^2)^{1/2}$, \\
%%$(R \times \hat{z}) \cdot \hat{n} = - (\partial_s R_z) (R \times \hat{z}) \cdot \hat{t}= - $
%and the total torque due to its translational motion is
Similarly, to total torque associated with the pure translation 
$\vect{u} = (0, 0, U)=U\hat{z}$ is:
\beq
L_z^{(Translation)} = U \int (C_n-C_t) (\partial_s R_z) R_{\perp} \cos \psi  ds \,.
\eeq
Decoupling the body motion into a pure rotation and a pure
translation, the $z$-component of total torque on the body is
expressed as
\beq
L_z = L_z^{(Translation)} + L_z^{(Rotation)} \,.
\eeq
In our experiments, we do not apply any external torque. Balancing two
torques gives an expression for the ratio of angular velocity to
translational velocity as
\beqs
&&L_z^{(Translation)} = - L_z^{(Rotation)} \nonumber \\
& \Rightarrow &
\f{\Omega}{U} = \f{\int (C_n-C_t) (\partial_s R_z) R_{\perp} \cos
\psi  ds}{\int R_{\perp}^2 \left[ C_t \cos^2 \psi + C_n \sin^2
\psi \right] ds} \,.
\eeqs
Note that this ratio of velocities depends only upon the ratio of drag
coefficients $\frac{C_n}{C_t}$.  Evaluating numerically these
integrals, and using $\vect{R}$ and $C_n = 2 C_t$ (the leading order
result of slender-body theory), $\Omega/U$ is plotted as squares in
Fig. \ref{Figure_transition}.  

The slope ($-B/D$) in the relation (\ref{eq:linear}) is measured in
experiments by using a least-squares fit of the data (see
Fig. \ref{Figure_linear}).  Although the resistive force theory
predictions of $\Omega/U$ show the same general trend as the
experimental measurements, the quantitative agreement is quite poor,
and the transition from clockwise to counter-clockwise rotation is not
captured.
%In fact,
%for larger values of RK, there is more than one hundred percent error in the predicted ratio.

%%%%%%%%%%%%%%%%%%%%%%%%%%%%%%%%%%%%%%%%%%%%%%%%%%%%%%%%%%%%%%%%%
%\begin{figure} %%result_test_Gray_measure_error.m
%    \centering
%        \includegraphics[width=.5\textwidth]{Fig_transition_error}
%\caption{Error is defined as the difference of two predicted values from different methods divided by the value from Regularized Stokeslet method. Filled and open circles are for Set I and II, respectively.}
%    \label{Figure_error}
%\end{figure}
%%%%%%%%%%%%%%%%%%%%%%%%%%%%%%%%%%%%%%%%%%%%%%%%%%%%%%%%%%%%%%%%%%

\section{Conclusion}

In conclusion, we have studied the rotational dynamics of superhelices
built out of superimposed helices of opposite handedness in a viscous
fluid. As the radius of the axial helix increases, we have observed
the transition of rotational direction to the natural direction of
rotation of the axial helix.  The rotational velocities corresponding
to translational velocities are also predicted semi-analytically using
resistive force theory, as well as numerically, using the method of
Regularized Stokeslets \cite {Cortez01, Cortez05}.  We see that
although there is qualitative agreement between experiments and
resistive force theory, for larger radii of the axial helix, the
predicted ratios of rotation to translational velocities differ from
experimental results by more than one hundred percent, and show the
incorrect direction of rotation.  In contrast, the easily implemented
computational framework of Regularized Stokeslets demonstrates
excellent quantitative agreement with experiments.

While this study is  motivated by the fascinating geometry of spirochetes,
we recognize that the rotation of the anterior superhelix of a spirochete is
not due to an imposed rotation about a vertical axis, 
but due to the counter-rotation of the periplasmic flagellum
(that determines the axial helix), against the cell body (the small helix).
The fluid mechanic implications of this 
counter-rotation 
that govern spirochete motility is examined in \cite {mcgf07}.

We thank J. Zhang, E. Kim, A. Medovikov, R. Cortez for helpful discussions.
This work was partially supported by  by the DOE (Grant No. DE-FG02- 88ER25053) and
NSF DMS 0201063.

%%%%%%%%%%%%%%%%%%%%%%%%%%%%%%%%%%%%%%%%%%%%%%%%%%%%%%%%%%%%%%%%
%\begin{figure} %%result_tes.m
%    \centering
%        \includegraphics[width=.5\textwidth]{Fig_transition_normal.eps}
%\caption{}
%    \label{Figure_normal}
%\end{figure}
%%%%%%%%%%%%%%%%%%%%%%%%%%%%%%%%%%%%%%%%%%%%%%%%%%%%%%%%%%%%%%%%%

%\section{I Need a help}
%{\bf Goal : Can we predict the transition point analytically?}


\begin{thebibliography}{99}
\bibitem{Taylor52} G. I. Taylor,
``The action of waving cylindrical tails in propelling microscopic
organisms,'' Proc. Royal Soc. London A {\bf 211}, 225--239 (1952).
\bibitem{Berg73}H. C. Berg,``Bacteria swim by
rotating their flagellar filaments,'' Nature {\bf 245}, 380--382
(1973).
%\bibitem{Berg00}H. C. Berg, Physics Today, January, 25--29
%(2000).
%\bibitem{Berg03}H. C. Berg, Annu. Rev. Biochem., {\bf 72}, 19--54
%(2003).
\bibitem{Lighthill76}J. Lighthill, ``Flagellar Hydrodynamics,'' SIAM
review {\bf 18}, 161--230 (1976).
\bibitem{Brownberg} D. Brown and H. C. Berg, ``Temporal stimulation of chemotaxis in Escherichia coli,'' Proc. Natl. Acad. Sci. U.S.A., {\bf 71}, 1388 (1974).
\bibitem{Macnab} R. M. MacNab and D. E. Koshland, ``The gradient-sensing mechanism in bacterial chemotaxis,'' Proc. Natl. Acad. Sci. U.S.A., {\bf 69}, 2509 (1972).
\bibitem{Purcell97}E. Purcell, ``The efficiency of propulsion by
a rotating flagellum,'' Proc. Natl. Acad. Sci. U.S.A. {\bf 94},
11307 (1997).
\bibitem{Powers} M. Kim, J. Bird, A. Van Parys, K. Breuer, T. Powers, ``A macroscopic scale model of bacterial flagellar bundling,'' Proc. Natl. Acad. Sci. U.S.A., {\bf 100},
15481-15485 (2003).
\bibitem{Flores} H. Flores, E. Lobaton, S. Mendez-Diez, S. Tlupova, R. Cortez, ``A study of baterial flagellar bundling ", Bull. Math. Biol., {\bf 67}, 137-168 (2005).
\bibitem{CDMF84}N. W. Charon, G. R. Daughtry, R. S. McCuskey, and
G. N. Franz, ``Microcinematographic analysis of Tethered {\em Leptospira illini}",
J. Bacter., {\bf 160} , 1067-1073 (1984).
\bibitem{Lepto} S. Goldstein and N. Charon, ``Multiple-exposure photographic analysis of a motile
spirochete", Proc. Natl. Acad. Sci. U.S.A., {\bf 87}, 4895-4899 (1990).
\bibitem{Cortez05}R. Cortez, L. Fauci, and A. Medovikov,
``The method of Regularized Stokeslets in three dimensions:
Analysis, validation, and application to helical swimming,'' Phys.
Fluids {\bf 17}, 031504 (2005).
\bibitem{Cortez01}R. Cortez, ``The method of Regularized Stokeslets,''
SIAM J. Sci. Comput. (USA) {\bf 23}, 1204 (2001).
\bibitem{Jung06}S. Jung, S. E. Spagnolie, K. Prikh, M. Shelley and A.-K. Tornberg
``Periodic sedimentation in a Stokesian fluid,'' Phys. Rev. E {\bf
74}, 035302(R) (2006).
\bibitem{Poz} C. Pozrikidis, ``Boundary integral and singularity methods for linearized viscous flow,'' Cambridge University
Press, Cambridge (1992).
%\bibitem{Levett01}P. N. Levett, Clin. Microbio. Rev., {\bf 14},
%296--326 (2001).
%\bibitem{BT79}H. C. Berg and L. Turner, Nature, {\bf 278}, 349--351 (1979).
\bibitem{Hancock53}G. J. Hancock, ``The self-propulsion of
microscopic organisms through liquids.'' Proc. R. Soc. Lond. A
{\bf 127}, 96--121 (1953).
\bibitem{Gray55}J. Gray and G. J. Hancock, ``The propulsion of
sea-urchin spermatozoa'' J. Exp. Biol. {\bf 32}, 802--814 (1955).
%\bibitem{Chwang71}A. T. Chwang and T. Y. Wu, ``A note on the
%helical movement of micro-organisms.'' Proc. R. Soc. Lond. B {\bf
%178}, 327--346 (1971).
\bibitem{mcgf07} A. Medovikov, R. Cortez, S. Goldstein, L. Fauci, ``Fluid mechanics of 
spirochete motility'', submitted (2007).



\end{thebibliography}
\end{document}